\documentstyle{article}
\newtheorem{rem}{Remark}

\begin{document}
\begin{center}
\Large{\bf  ON THE RIEMANN EXTENSION OF ROTATING \\[2mm] MINKOWSKY SPACE-TIME
METRIC}\vspace{4mm}\normalsize
\end{center}
 \begin{center}
\Large{\bf Valery Dryuma}\vspace{4mm}\normalsize
\end{center}
\begin{center}
{\bf Institute of Mathematics and Informatics AS Moldova, Kishinev}\vspace{4mm}\normalsize
\end{center}
\begin{center}
{\bf  E-mail: valery@dryuma.com;\quad cainar@mail.md \quad dryuma@math.md}\vspace{4mm}\normalsize
\end{center}
\begin{center}
{\bf  Abstract}\vspace{4mm}\normalsize
\end{center}

   Some properties of the Riemann extension of Minkowsky space-time metric in
   rotating coordinate system are studied.

\section{Introduction}

   The notion of  Riemann extension of nonriemannian spaces was introduced first in
    ~(\cite{dryuma:paterson&walker}).
Main idea of this theory is application of the methods of Riemann
geometry for studying of the properties of nonriemaniann spaces.

For example the system differential equations in form
\begin{equation} \label{dryuma:eq1}
\frac{d^2 x^{k}}{ds^2}+\Pi^k_{ij}\frac{dx^{i}}{ds}\frac{dx^{j}}{ds}=0
\end{equation}
with arbitrary coefficients $\Pi^k_{ij}(x^l)$ can be considered as the system of geodesic equations of
affinely connected space with local coordinates $x^k$.

 For the n-dimensional Riemannian spaces with the metrics
\[
{^n}ds^2=g_{ij}dx^i dx^j
\] the system of geodesic equations looks
similar but the coefficients $\Pi^k_{ij}(x^l)$ now have very special form and depend
from the choice of the metric $g_{ij}$.
\[
\Pi^i_{kl}=\Gamma^i_{kl}=\frac{1}{2}g^{im}(g_{mk,l}+g_{ml,k}-g_{kl,m})
\]

In order that methods of Riemann geometry can be applied for studying of the properties of the spaces with
equations (\ref{dryuma:eq1}) the construction of 2n-dimensional extension of the space with local coordinates $x^i$  was introduced .

The metric of extended space constructs with help of coefficients of equation (\ref{dryuma:eq1}) and looks as follows
\begin{equation} \label{dryuma:eq2}
{^{2n}}ds^2=-2\Pi^k_{ij}(x^l)\Psi_k dx^i dx^j+2d \Psi_k dx^k
\end{equation}
where $\Psi_{k}$ are the coordinates of additional space.

   The important property of such type metric is that the geodesic
 equations of metric (\ref{dryuma:eq2})  decomposes into the two parts
\begin{equation} \label{dryuma:eq3}
\ddot x^k +\Gamma^k_{ij}\dot x^i \dot x^j=0,
\end{equation}
and
\begin{equation} \label{dryuma:eq4}
\frac{\delta^2 \Psi_k}{ds^2}+R^l_{kji}\dot x^j \dot x^i \Psi_l=0,
\end{equation}
where
\[
\frac{\delta \Psi_k}{ds}=\frac{d \Psi_k}{ds}-\Gamma^l_{jk}\Psi_l\frac{d x^j}{ds}.
\]

The first part (\ref{dryuma:eq3}) of full system is the system of equations for
geodesics of basic space with local coordinates $x^i$ and it does not contains the
coordinates $\Psi_k$.

 The second part (\ref{dryuma:eq4}) of system of geodesic equations  has the form
of linear $4\times4$ matrix system of second order ODE's for coordinates $\Psi_k$
\begin{equation} \label{dryuma:eq5}
\frac{d^2 \vec \Psi}{ds^2}+A(s)\frac{d \vec \Psi}{ds}+B(s)\vec \Psi=0
\end{equation}
with the matrix
\[
A(s)=A(x^i(s),\dot x^i(s)), \quad B(s)=B(x^i(s),\dot x^i(s)).
\]

From this point of view we have the case of geodesic extension of
the basic space  $(x^i)$.

  It is important to note that the geometry of extended space is depended from
geometry of basic space.

   For example the property of such type of the space to be a Ricci-flat
keeps also for the extended space.

     This fact give us the possibility to use the linear system of equation (\ref{dryuma:eq5}) for studying
of the properties of basic space.

  In particular the invariants of $4\times4$
matrix-function
\[
E=B-\frac{1}{2}\frac{d A}{ds}-\frac{1}{4}A^2
\]
under change of the coordinates $\Psi_k$ can be of used for that.

   Remark that for extended spaces all scalar invariants constructed with
   the help of curvature tensor and its covariant derivatives are vanishing.

   The first applications of the notion of extended spaces to the studying of nonlinear second order differential
 equations connected with nonlinear dynamical systems were done in works of author
  ~(\cite{dryuma1:dryuma,dryuma2:dryuma,dryuma3:dryuma,dryuma4:dryuma}).

 Here we consider the properties of Riemann extension of the Minkowsky space in
  rotating system  of coordinates.

\section{The  geodesic equations of rotating coordinate system}.

The line element of standard metric of the Minkowsky space-time in rotating
coordinate system $x ,\phi,z,t$ has the form

\begin{equation} \label{dryuma:eq6}
ds^2=-dx^2-x^2d\phi^2-dz^2-2\Omega x^2 d\phi d t+(c^2-\Omega^2x^2) dt^2.
\end{equation}

   Here the parameters $\Omega$ and $c$ are the velocity of rotation
   and the velocity of light.

The geodesic equations of the metric ~(\ref{dryuma:eq6}) are given by
\begin{equation} \label{dryuma:eq7}
{\frac {d^{2}}{d{s}^{2}}}x \left( s \right) -x \left( s \right)
 \left( {\frac {d}{ds}}\phi \left( s \right)  \right) ^{2}-2\,\Omega\,
x \left( s \right)  \left( {\frac {d}{ds}}\phi \left( s \right)
 \right) {\frac {d}{ds}}t \left( s \right) -{\Omega}^{2}x \left( s
 \right)  \left( {\frac {d}{ds}}t \left( s \right)  \right) ^{2}=0,
\end{equation}
\begin{equation} \label{dryuma:eq8}
 \left( {\frac {d^{2}}{d{s}^{2}}}\phi \left( s \right)  \right) x
 \left( s \right) +2\, \left( {\frac {d}{ds}}x \left( s \right)
 \right) {\frac {d}{ds}}\phi \left( s \right) +2\,\Omega\, \left( {
\frac {d}{ds}}x \left( s \right)  \right) {\frac {d}{ds}}t \left( s
 \right)=0,
\end{equation}
\begin{equation}\label{dryuma:eq9}
{\frac {d^{2}}{d{s}^{2}}}z(s)=0,
\end{equation}
\begin{equation} \label{dryuma:eq10}
{\frac {d^{2}}{d{s}^{2}}}t(s)=0.
\end{equation}

The symbols of Christoffel of the metric are
$$
\Gamma^1_{22}=-x,\quad
\Gamma^1_{24}=-\Omega x,\quad \Gamma^1_{44}=-\Omega^2 x,
\quad \Gamma^2_{12}=\frac{1}{x},\quad
\Gamma^2_{14}=\frac{\Omega}{x}.
$$

    The equations of geodesic  (\ref{dryuma:eq7}--\ref{dryuma:eq10}) have
     the solutions
\begin{equation}\label{dryuma:eq11}
x(s)={\frac {{A}}{\sqrt {{\frac {d}{ds}}\phi(s)+\Omega\,a}}}, \quad
\phi(s)=-\Omega\,as+\arctan(Bs+C)+E,
\end{equation}
\begin{equation}\label{dryuma:eq12}
 \dot t=a,
\quad \dot z =b.
\end{equation}
where a $point$ denotes differentiation with respect to parameter $s$ and
 $(A,B,C,E,a,b)$ are the constants of motion.

\section{ Eight-dimensional extension of metric of rotating space-time}

        Now with help of the formulae (\ref{dryuma:eq2}) we construct the eight-dimensional
extension of basic metric (\ref{dryuma:eq6}).

 It has the form
\begin{equation}\label{dryuma:eq13}
^{8}ds^2=-\frac{4}{x} Q d x d \phi-\frac{4 \Omega}{x}Q dx dt+2 x P d \phi^2+4 x
\Omega P d \phi dt+2x \Omega^2 P d t^2+ \]\[+2dx dP+2d \phi d Q +2 d z dU +2 dt d V,
\end{equation}
where $(P,Q,U,V)$ are an additional coordinates.

    The eight-dimensional space in local coordinates $(x,\phi,z, t,P,Q,U,V)$
 with the metric (\ref{dryuma:eq13}) is also a flat space.
 Its Riemann tensor equal to zero
\[
{^8}R_{iklm}=0.
\]

  Full system of  geodesic equations for the metric (\ref{dryuma:eq7})
   decomposes into the two parts.

The first part coincides with the equations (\ref{dryuma:eq7}-\ref{dryuma:eq10})
on the coordinates $(x,\phi,z, t)$ and second  part forms the linear system of equations for coordinates
$P,Q,U,V$.

 They are defined as
\begin{equation}\label{dryuma:eq14}
{\frac {d^{2}}{d{s}^{2}}}P \left( s \right) -{\frac { \left( 2\,\Omega \, \left(
{\frac {d}{ds}}t \left( s \right)  \right) x+2\, \left( { \frac {d}{ds}}\phi \left(
s \right)  \right) x \right) {\frac {d}{ds}} Q \left( s \right)
}{{x}^{2}}}-\]\[-{\frac { \left( 2\,\Omega\, \left( { \frac {d}{ds}}\phi \left( s
\right)  \right)  \left( {\frac {d}{ds}}t
 \left( s \right)  \right) {x}^{2}+{\Omega}^{2} \left( {\frac {d}{ds}}
t \left( s \right)  \right) ^{2}{x}^{2}+ \left( {\frac {d}{ds}}\phi
 \left( s \right)  \right) ^{2}{x}^{2} \right) P \left( s \right) }{{x
}^{2}}}+\]\[+{\frac { \left( 4\, \left( {\frac {d}{ds}}x \left( s \right)
 \right) {\frac {d}{ds}}\phi \left( s \right) +4\,\Omega\, \left( {
\frac {d}{ds}}x \left( s \right)  \right) {\frac {d}{ds}}t \left( s
 \right)  \right) Q \left( s \right) }{{x}^{2}}}=0,
\end{equation}
\\[2mm]
\begin{equation}\label{dryuma:eq15}
{\frac {d^{2}}{d{s}^{2}}}Q \left( s \right) -2\,{\frac { \left( { \frac {d}{ds}}x
\left( s \right)  \right) {\frac {d}{ds}}Q \left( s
 \right) }{x}}-2\,{\frac { \left( -{x}^{3}{\frac {d}{ds}}\phi \left( s
 \right) -{x}^{3}\Omega\,{\frac {d}{ds}}t \left( s \right)  \right) {
\frac {d}{ds}}P \left( s \right) }{{x}^{2}}}-\]\[-2\,{\frac { \left( -
 \left( {\frac {d}{ds}}x \left( s \right)  \right) ^{2}+ \left( {
\frac {d}{ds}}\phi \left( s \right)  \right) ^{2}{x}^{2}+2\,\Omega\,
 \left( {\frac {d}{ds}}\phi \left( s \right)  \right)  \left( {\frac {
d}{ds}}t \left( s \right)  \right) {x}^{2}+{\Omega}^{2} \left( {\frac {d}{ds}}t
\left( s \right)  \right) ^{2}{x}^{2} \right) Q \left( s
 \right) }{{x}^{2}}}-\]\[-{\frac { \left(  \left( {\frac {d}{ds}}x \left( s
 \right)  \right)  \left( {\frac {d}{ds}}\phi \left( s \right)
 \right) {x}^{2}+\Omega\, \left( {\frac {d}{ds}}x \left( s \right)
 \right)  \left( {\frac {d}{ds}}t \left( s \right)  \right) {x}^{2}
 \right) P \left( s \right) }{{x}^{2}}}=0,
\end{equation}
\\[2mm]
\begin{equation}\label{dryuma:eq16}
{\frac {d^{2}}{d{s}^{2}}}U(s)=0,
\end{equation}
\\[2mm]
\begin{equation}\label{dryuma:eq17}
{\frac {d^{2}}{d{s}^{2}}}V \left( s \right) -2\,{\frac {\Omega\,
 \left( {\frac {d}{ds}}x \left( s \right)  \right) {\frac {d}{ds}}Q
 \left( s \right) }{x}}-2\,{\frac {\Omega\, \left( -{x}^{3}{\frac {d}{
ds}}\phi \left( s \right) -{x}^{3}\Omega\,{\frac {d}{ds}}t \left( s
 \right)  \right) {\frac {d}{ds}}P \left( s \right) }{{x}^{2}}}-\]\[-2\,{
\frac {\Omega\, \left( - \left( {\frac {d}{ds}}x \left( s \right)
 \right) ^{2}+ \left( {\frac {d}{ds}}\phi \left( s \right)  \right) ^{
2}{x}^{2}+2\,\Omega\, \left( {\frac {d}{ds}}\phi \left( s \right)
 \right)  \left( {\frac {d}{ds}}t \left( s \right)  \right) {x}^{2}+{
\Omega}^{2} \left( {\frac {d}{ds}}t \left( s \right)  \right) ^{2}{x}^ {2} \right) Q
\left( s \right) }{{x}^{2}}}+\]\[+2\,{\frac {\Omega\, \left(
 \left( {\frac {d}{ds}}x \left( s \right)  \right)  \left( {\frac {d}{
ds}}\phi \left( s \right)  \right) {x}^{2}+\Omega\, \left( {\frac {d}{ ds}}x \left(
s \right)  \right)  \left( {\frac {d}{ds}}t \left( s
 \right)  \right) {x}^{2} \right) P \left( s \right) }{{x}^{2}}}=0.
\end{equation}

     In result we have got a linear  matrix-second order ODE for the coordinates $U,V,P,Q$
\begin{equation}\label{dryuma:eq18}
\frac{d^2\Psi}{ds^2}+A(x,\phi,z,t)\frac{d\Psi}{ds}+B(x,\phi,z,t)\Psi=0,
\end{equation}
where
\[
\Psi(s)=\left(\begin{array}{c}
P(s)\\
Q(s)\\
U(s)\\
V(s)
\end{array}\right)
\]
and $A,B$ are some $4 \times 4$ matrix-functions depending from the coordinates $x(s),\phi(s),z(s),t(s)$
and their derivatives.

    Now we shall investigate the properties of the matrix system of equations
     (\ref{dryuma:eq14}-\ref{dryuma:eq17}).

    To integrate this system we use the relation
\begin{equation}\label{dryuma:eq19}
\dot x(s) P(s)+\dot \phi(s) Q(s)+\dot z(s) U(s)+ \dot t(s) V(s)-\frac{s}{2}=const,
\end{equation}
which is valid for the every Riemann extensions of affinely connected space and
which is consequence of the well known first integral of geodesic equations of
arbitrary Riemann space
\[
g_{ik}\dot x^{i}\dot x^{k}=const.
\]

    Using a following solutions for the coordinates $U(s), t(s), z(s)$
\begin{equation}\label{dryuma:eq20}
U(s)=us+l,\quad t=as+m,\quad z=bs+n
\end{equation}
 we get
\begin{equation}\label{dryuma:eq21}
V(s)=-1/2\,{\frac {2\,\left ({\frac {d}{ds}}x(s)\right )P(s)+2\,\left ({
\frac {d}{ds}}\phi(s)\right )Q(s)+2\,bus-s}{a}}.
\end{equation}

  After substitution given expression into the equation for coordinate $V(s)$)
  one get the identity.

   So our problem is reduced to integration of the system of  second order ODE's
 (\ref{dryuma:eq14},\ref{dryuma:eq15}).

  Further we use a solutions for coordinates $x(s)$ and $\phi(s)$ in form
\begin{equation}\label{dryuma:eq22}
x(s)={\frac {A}{\sqrt { \left( {s}^{2}+1 \right) ^{-1}}}}
\end{equation}
and
\begin{equation}\label{dryuma:eq23}
\phi(s)=-\Omega\,as+\arctan(s).
\end{equation}

   At this condition the system (\ref{dryuma:eq14}-\ref{dryuma:eq17}) takes a form
\[
{\frac {d^{2}}{d{s}^{2}}}P \left( s \right) ={\frac { \left( 2\,{s}^{2 }+2 \right)
{\frac {d}{ds}}Q \left( s \right) }{\sqrt {{s}^{2}+1}
 \left( {s}^{4}+2\,{s}^{2}+1 \right) }}+{\frac {P \left( s \right) }{{
s}^{4}+2\,{s}^{2}+1}}-4\,{\frac {Q \left( s \right) s}{\sqrt {{s}^{2}+ 1} \left(
{s}^{4}+2\,{s}^{2}+1 \right) }},
\]

 \[
{\frac {d^{2}}{d{s}^{2}}}Q \left( s \right) ={\frac { \left( 2\,{s}^{2 }+2 \right)
{\frac {d}{ds}}Q \left( s \right) }{\sqrt {{s}^{2}+1}
 \left( {s}^{4}+2\,{s}^{2}+1 \right) }}+{\frac {P \left( s \right) }{{
s}^{4}+2\,{s}^{2}+1}}-4\,{\frac {Q \left( s \right) s}{\sqrt {{s}^{2}+ 1} \left(
{s}^{4}+2\,{s}^{2}+1 \right) }}.
 \]

   Its solutions is given by
\begin{equation}\label{dryuma:eq24}
P \left( s \right) ={\frac {{C_2}+{C_3}\,s+{ C_4}\,{s}^{ 2}}{\sqrt {{s}^{2}+1}}} ,
\end{equation}
and
\begin{equation}\label{dryuma:eq25}
Q(s)={ C_1}\, \left( {s}^{2}+1 \right) +s \left( {C_4}-{ C_2}
 \right) +{ C_3},
\end{equation}
where $C_i$ are parameters.

Using these expressions we find from the relation ~(\ref{dryuma:eq21})
\begin{equation}\label{dryuma:eq26}
V \left( s \right) =\Omega\,{s}^{2}{C_1}+1/2\,{\frac { \left( -2 \,{
C_4}-2\,\Omega\,a{ C_2}-2\,bu+1+2\,\Omega\,a{ C_4}
 \right) s}{a}}+\]\[+1/2\,{\frac {-2\,{C_1}-2\,{C_3}+2\,\Omega\,a
{ C_3}+2\,\Omega\,a{ C_1}}{a}}.
\end{equation}

 So the formulaes ~(\ref{dryuma:eq20}),~(\ref{dryuma:eq24}-\ref{dryuma:eq26}),
 represent the solutions of geodesic equations of the Riemann extension of the
 Minkowsky space-time in rotating coordinate system.

   To the geodesic motion in a basic space corresponds the geodesic motion in
   extended space (partner space).

    A studying of the properties of both types of motions may be useful
    from various points of view.

\section{Three-dimensional spatial metric
of rotating flat space-time}

   The properties of three-dimensional subspace of rotating coordinate system of the Minkowsky
   space-time are determined by the metric
\[
^{3}ds^2=dr^2+dz^2+\frac{r^2}{1-\Omega r^2/c^2}d\phi^2
\]
or
\begin{equation}\label{dryuma:eq27}
 {{\it ds}}^{2}= d{x}^2
+{x}^{2} d{y}^2\left (1-{\frac {{\Omega}^
{2}{x}^{2}}{{c}^{2}}}\right )^{-1}+dz^2.
\end{equation}

 The Ricci tensor in this case is
\[
R_{a b}=\left [\begin {array} {ccc} -3\,{\frac
{{c}^{2}{\Omega}^{2}}{\left ({c}^{2}-{\Omega}^{2}{x}^ {2}\right
)^{2}}}&0&0\\\noalign{\medskip}0&-3\,{\frac {{c}^{4}{x}^{2}{
\Omega}^{2}}{\left ({c}^{2}-{\Omega}^{2}{x}^{2}\right )^{3}}}&0
\\\noalign{\medskip}0&0&0\end {array}\right ].
\]

   A simplest scalar invariant
\[
K=R^{ij}R_{i j}=18\,{\frac {{c}^{4}{\Omega}^{4}}{\left
(-{c}^{2}+{\Omega}^{2}{x}^{2} \right )^{4}}}
\]
has  singularity at the $x=\frac{c}{\Omega}$.

\begin{rem}

 In theory of varieties the Chern-Simons characteristic class is constructed from a
 matrix gauge connection $A^i_{jk}$ as
\[
W(A)=\frac{1}{4\pi^2}\int d^3x
\epsilon^{ijk}tr\left(\frac{1}{2}A_i\partial_j
A_k+\frac{1}{3}A_iA_jA_k \right).
\]

   This term can be translated into three-dimensional geometric quantity by replacing the matrix connection
   $A^i_{jk}$ with the Christoffel connection $\Gamma^i_{jk}$.

   For the density of  Chern-Simons invariant the
   expression (\cite{dryuma:jac}) can be obtained
\begin{equation} \label{dryuma:eq28}
CS(\Gamma)=\epsilon^{i j k}(\Gamma^p_{i q}\Gamma^q_{k
p;j}+\frac{2}{3}\Gamma^p_{i q}\Gamma^q_{j r}\Gamma^r_{k p}).
\end{equation}

   For the metric (\ref{dryuma:eq27}) this  quantity is
\[
CS(\Gamma)=0.
\]

 The variation of $W(\Gamma)$ with respect to metric $g_{i j}$
 can be presented as
\[
\delta W(\Gamma)=-\frac{1}{4 \pi^2} \int d^3 x \delta
g_{\mu\nu}\sqrt{g}C^{\mu\nu},
\]
where
\[
C^{\mu\nu}=\frac{1}{2 \sqrt{g}}\left(\epsilon^{\mu i j}
D_{i}R^{\mu}_j+\epsilon^{\nu i j} D_{j}R^{\nu}_i\right)
\]
is the Cotton tensor which  has an important role in 3-dimensional
geometry.

$C^{\mu\nu}$ vanishes if and only if the space is conformally flat
\[
C^{\mu\nu}=0 \Longleftrightarrow conformally \quad flat \quad
space.
\]

In case of the metric ~9\ref{dryuma:eq27}) we get
$C^{\mu\nu}\not=0$
\[
2{\frac {xc}{\sqrt {{c}^{2}-{\Omega}^{2}{x}^{2}}}}
C^{\mu\nu}=\left [\begin {array}{ccc}
0&0&0\\\noalign{\medskip}0&0&12\,{\frac {c{ \Omega}^{4}}{\left
({c}^{2}-{\Omega}^{2}{x}^{2}\right )^{2}\sqrt {-{c}
^{2}+{\Omega}^{2}{x}^{2}}}}\\\noalign{\medskip}0&12\,{\frac
{c{\Omega} ^{4}}{\left ({c}^{2}-{\Omega}^{2}{x}^{2}\right
)^{2}\sqrt {-{c}^{2}+{ \Omega}^{2}{x}^{2}}}}&0\end {array}\right ]
\]
and corresponding space is not conformally flat.
\end{rem}

  The properties of the metric (\ref{dryuma:eq27})
     can be studied with  help of the eigenvalue equation for the Laplace-de Rham operator defined on the
     1-forms
\[
A(x,y,z)=A_i(x,y,z)dx^i.
\]

   It is given the by expression
\[
g^{ij}\nabla_i\nabla_j A_k-R^l_k A_l+\lambda A_k=0.
\]

    In considered case these equations
    take a form ($A_i=[u(x,y,z),v(x,y,z),w(x,y,z)]$)
\[
{\frac {\partial ^{2}}{\partial {x}^{2}}}u(x,y,z)+{\frac {\left
({c}^{ 2}-{\Omega}^{2}{x}^{2}\right ){\frac {\partial
^{2}}{\partial {y}^{2}} }u(x,y,z)}{{c}^{2}{x}^{2}}}+{\frac
{\partial ^{2}}{\partial {z}^{2}}}u (x,y,z)+{\frac {{c}^{2}{\frac
{\partial }{\partial x}}u(x,y,z)}{\left
({c}^{2}-{\Omega}^{2}{x}^{2}\right )x}}-2\,{\frac {{\frac
{\partial }{
\partial y}}v(x,y,z)}{{x}^{3}}}+\]\[+\left (\lambda-{\frac {{c}^{2}\left ({
c}^{2}-3\,{\Omega}^{2}{x}^{2}\right )}{{x}^{2}\left (c-\Omega\,x
\right )^{2}\left (c+\Omega\,x\right )^{2}}}\right )u(x,y,z) =0,
\]
\[
{\frac {\partial ^{2}}{\partial {x}^{2}}}v(x,y,z)+{\frac {\left
({c}^{ 2}-{\Omega}^{2}{x}^{2}\right ){\frac {\partial
^{2}}{\partial {y}^{2}} }v(x,y,z)}{{c}^{2}{x}^{2}}}+{\frac
{\partial ^{2}}{\partial {z}^{2}}}v (x,y,z)-{\frac {\left ({\frac
{\partial }{\partial x}}v(x,y,z)\right ) {c}^{2}}{\left
({c}^{2}-{\Omega}^{2}{x}^{2}\right )x}}+2\,{\frac {{c}^ {2}{\frac
{\partial }{\partial y}}u(x,y,z)}{\left ({c}^{2}-{\Omega}^{2
}{x}^{2}\right )x}}+\]\[+\lambda\,v(x,y,z)  =0,
\]
\[
{\frac {\partial ^{2}}{\partial {x}^{2}}}w(x,y,z)+{\frac {\left
({c}^{ 2}-{\Omega}^{2}{x}^{2}\right ){\frac {\partial
^{2}}{\partial {y}^{2}} }w(x,y,z)}{{c}^{2}{x}^{2}}}+{\frac
{\partial ^{2}}{\partial {z}^{2}}}w (x,y,z)+{\frac {\left ({\frac
{\partial }{\partial x}}w(x,y,z)\right ) {c}^{2}}{\left
({c}^{2}-{\Omega}^{2}{x}^{2}\right )x}}+\]\[+\lambda\,w(x,y, z)
=0.
\]

    The solutions of given system depend from the eigenvalues $\lambda$ and
    characterize the topological  properties of the metric.

   Now we shall investigate the properties of geodesics of three-dimensional
   spatial geometry.

   They are defined by the system of equations
\begin{equation}\label{dryuma:eq28}
{\frac {d^{2}}{d{s}^{2}}}x(s)-{\frac {x(s){c}^{4}\left ({\frac
{d}{ds} }y(s)\right )^{2}}{\left (\Omega\,x(s)-c\right )^{2}\left
(\Omega\,x(s )+c\right )^{2}}}=0 ,
\end{equation}
\begin{equation}\label{dryuma:eq29}
{\frac {d^{2}}{d{s}^{2}}}y(s)-2\,{\frac {{c}^{2}\left ({\frac
{d}{ds}} x(s)\right ){\frac {d}{ds}}y(s)}{x(s)\left
(\Omega\,x(s)-c\right ) \left (\Omega\,x(s)+c\right )}}=0 ,
\end{equation}
\begin{equation}\label{dryuma:eq30}
{\frac {d^{2}}{d{s}^{2}}}z(s)=0, \end{equation}
 where the last equation is independent.

  The system (\ref{dryuma:eq28})-(\ref{dryuma:eq29}) has the first integral
\[
\left ({\frac {d}{ds}}x(s)\right )^{2}+\left (x(s)\right
)^{2}\left ({ \frac {d}{ds}}y(s)\right )^{2}\left (1-{\frac
{{\omega}^{2}\left (x(s) \right )^{2}}{{c}^{2}}}\right
)^{-1}+\mu=0.
\]

        From the second equation of the system  we find
\[
y(s)={C_1}+{C_2}\,\int \!{\frac {\left (\Omega\,x(s)-c \right
)\left (\Omega\,x(s)+c\right )}{\left (x(s)\right )^{2}}}{ds}.
\]

  After substitution this expression into the first equation it takes the form
\[
\left ({\frac {d^{2}}{d{s}^{2}}}x(s)\right )\left (x(s)\right
)^{3}-{c }^{4}{C_2}^{2}=0.
\]

    Its solution is
\[
x(s)={\frac {\sqrt {{C_3}\,\left
({c}^{4}{{C_2}}^{2}+{s}^{2}{{C_3}}^{2}+2\,s{{C_3}}^{2}{C_4}+{{C_4}}^{2}{{C_3}}^{2}\right
)}}{{C_3}}} .
\]

    Taking in  consideration the first integral we find
\[
C_3={c}^{2}{{C_2}}^{2}{\Omega}^{2}+\mu.
\]

    Using this relation we get
\[
x(s)^2=\left ({c}^{2}{{C_2}}^{2}{\Omega}^{2}+\mu\right
){s}^{2}+2\, \left ({c}^{2}{{C_2}}^{2}{\Omega}^{2}+\mu\right
){C_4}\,s+\]\[+{ \frac
{{c}^{4}{{C_2}}^{2}+{{C_4}}^{2}{\mu}^{2}+{{C_4}}^
{2}{c}^{4}{{C_2}}^{4}{\Omega}^{4}+
2\,{{C_4}}^{2}{c}^{2}{{C_2}}^{2}{\Omega}^{2}\mu}{{c}^{2}{{C_2}}^{2}{\Omega}^{2}+\mu
}} ,
\]
and the equation for the function $y(s)$
\begin{equation}\label{dryuma:eq31}
{\frac {d}{ds}}y(s)-{\frac {A(s)}{B(s)}}=0,
\end{equation}
 where
\[
A(s)={ C_2}\,\left ({\Omega}^{2}{\mu}^{2}+2\,{c}^{2}\mu\,{{
C_2}}^{2}{\Omega}^{4}+{c}^{4}{{ C_2}}^{4}{\Omega}^{6}\right ){s}^
{2}+{C_2}\,\left (2\,{\mu}^{2}{\Omega}^{2}{ C_4}+4\,{c}^{2}
\mu\,{{ C_2}}^{2}{\Omega}^{4}{ C_4}+2\,{c}^{4}{{ C_2}}^{4}
{\Omega}^{6}{ C_4}\right )s+\]\[+{ C_2}\,\left
({\mu}^{2}{\Omega}^{ 2}{{ C_4}}^{2}+2\,{c}^{2}{{
C_2}}^{2}\mu\,{\Omega}^{4}{{ C_4}}^{2}+{\Omega}^{6}{{
C_4}}^{2}{c}^{4}{{ C_2}}^{4}-{c}^{2} \mu\right ),
\]
and
\[
B(s)=\left (2\,{c}^{2}{{C_2}}^{2}{\Omega}^{2}\mu+{\mu}^{2}+{c}^{4
}{{ C_2}}^{4}{\Omega}^{4}\right ){s}^{2}+\left (2\,{ C_4}\,{
\mu}^{2}+4\,{C_4}\,{c}^{2}{{ C_2}}^{2}{\Omega}^{2}\mu+2\,{
C_4}\,{c}^{4}{{ C_2}}^{4}{\Omega}^{4}\right )s+\]\[+{c}^{4}{{
C_2}}^{2}+{{ C_4}}^{2}{\mu}^{2}+{{ C_4}}^{2}{c}^{4}{{ C_2
}}^{4}{\Omega}^{4}+2\,{{ C_4}}^{2}{c}^{2}{{ C_2}}^{2}{\Omega}^
{2}\mu.
\]

     Explicit form of the function $y(s)$ depends  from relations between
     the parameters $(C_i, \mu)$ and may be obtained from  integration of the equation
     (\ref{dryuma:eq31}).

\begin{rem}

    The system of equations for $x(s)$ and $y(s)$ is equivalent to the equation
    on the function $y(x)$
\[
{\frac {d^{2}}{d{x}^{2}}}y \left( x \right) +2\,{\frac {{c}^{2}{\frac {d}{dx}}y
\left( x \right) }{x \left( {c}^{2}-{\Omega}^{2}{x}^{2}
 \right) }}+{\frac {{c}^{4}x \left( {\frac {d}{dx}}y \left( x \right)
 \right) ^{3}}{ \left( {c}^{2}-{\Omega}^{2}{x}^{2} \right) ^{2}}}=0.
\]

 Its first integral is defined by
 \[
{\frac {d}{dx}}y(x)=\nu\,\left ({c}^{2}-{\Omega}^{2}{x}^{2}\right
){x} ^{-2}{\frac {1}{\sqrt {1+{\frac
{{\Omega}^{2}{\nu}^{2}}{{c}^{2}}}-{ \frac
{{\nu}^{2}}{{x}^{2}}}}}}{c}^{-2} ,
 \]
where $\nu\neq 0$ is parameter.
\end{rem}

\section{Six-dimensional Riemann extension of spatial geometry}

    Now we shall study the Riemann extension of the metric (\ref{dryuma:eq27}).

    Using the corresponding coefficient of connection
\[
\Gamma{^y_{xy}}= {\frac {{c}^{2}}{x\left
({c}^{2}-{\Omega}^{2}{x}^{2}\right )}}, \quad
\Gamma{^x_{yy}}=-{\frac {x{c}^{4}}{\left
({c}^{2}-{\Omega}^{2}{x}^{2}\right )^{2}}}
\]
    we get the six-dimensional metric
\begin{equation}\label{dryuma:eq32}
^{6}ds^2=-4\,{\frac {{c}^{2}Vd{{x}}d{{y}}}{x\left
({c}^{2}-{\Omega}^{2}{x}^{2 }\right )}}+2\,{\frac
{x{c}^{4}U{d{{y}}}^{2}}{\left ({c}^{2}-{\Omega} ^{2}{x}^{2}\right
)^{2}}}+2\,d{{x}}d{{U}}+2\,d{{y}}d{{V}}+2\,d{{z }}d{{W}}
\end{equation}
for the space $^{6}D$ in local coordinates $(x,y,z,U,V,W)$.

   The Ricci tensor of the metric is
\[
\mbox {{R}}_{{a}}\mbox {{}}_{{b}}= \left[ \begin {array} {cccccc} -6\,{\frac
{{c}^{2}{\omega}^{2}}{ \left( -{c}^{2}+{\omega}^{2 }{x}^{2} \right)
^{2}}}&0&0&0&0&0\\\noalign{\medskip}0&-6\,{\frac {{x} ^{2}{c}^{4}{\omega}^{2}}{
\left( -{c}^{2}+{\omega}^{2}{x}^{2} \right)
^{3}}}&0&0&0&0\\\noalign{\medskip}0&0&0&0&0&0\\\noalign{\medskip}0&0&0
&0&0&0\\\noalign{\medskip}0&0&0&0&0&0\\\noalign{\medskip}0&0&0&0&0&0
\end {array} \right]
\]

    The geodesics of metric (\ref{dryuma:eq32}) is defined by the equations
    (\ref{dryuma:eq28})-(\ref{dryuma:eq30}) for the coordinates
    $x(s),y(s),z(s)$ and with help of the system of equations
\[
{\frac {d^{2}}{d{s}^{2}}}U(s)+2\,{\frac {{c}^{2}\left ({\frac
{d}{ds}} y(s)\right ){\frac {d}{ds}}V(s)}{x\left
(-{c}^{2}+{\Omega}^{2}{x}^{2} \right )}}+4\,{\frac {\left ({\frac
{d}{ds}}x(s)\right )V\left ({ \frac {d}{ds}}y(s)\right
){c}^{4}}{\left (\Omega\,x-c\right )^{2} \left (\Omega\,x+c\right
)^{2}{x}^{2}}}+{\frac {{c}^{4}\left ({\frac { d}{ds}}y(s)\right
)^{2}\left ({c}^{2}+3\,{\Omega}^{2}{x}^{2}\right )U} {\left
(\Omega\,x-c\right )^{3}\left (\Omega\,x+c\right )^{3}}} =0,
\]
\[
{\frac {d^{2}}{d{s}^{2}}}V(s)+2\,{\frac {{c}^{2}\left ({\frac
{d}{ds}} x(s)\right ){\frac {d}{ds}}V(s)}{x\left
(-{c}^{2}+{\Omega}^{2}{x}^{2} \right )}}+2\,{\frac {{c}^{4}\left
({\frac {d}{ds}}y(s)\right )x{ \frac {d}{ds}}U(s)}{\left
(\Omega\,x-c\right )^{2}\left (\Omega\,x+c \right
)^{2}}}-2\,{\frac {{c}^{4}\left ({\frac {d}{ds}}x(s)\right ) \left
({\frac {d}{ds}}y(s)\right )\left (-{c}^{2}+3\,{\Omega}^{2}{x}^{
2}\right )U(s)}{\left (\Omega\,x-c\right )^{3}\left (\Omega\,x+c
\right )^{3}}}-\]\[-2\,{\frac {{c}^{2}\left ({c}^{4}\left ({\frac
{d}{ds}}x (s)\right )^{2}-4\,{c}^{2}\left ({\frac
{d}{ds}}x(s)\right )^{2}{ \Omega}^{2}{x}^{2}+3\,\left ({\frac
{d}{ds}}x(s)\right )^{2}{\Omega}^{ 4}{x}^{4}-{c}^{4}\left ({\frac
{d}{ds}}y(s)\right )^{2}{x}^{2}\right ) V(s)}{{x}^{2}\left
(\Omega\,x-c\right )^{3}\left (\Omega\,x+c\right )^ {3}}} =0,
\]
\[
{\frac {d^{2}}{d{s}^{2}}}W(s)=0
\]
for additional coordinates $U,V,W$ of partner space.

     The first integral of geodesics of the metric (\ref{dryuma:eq32}) is defined by
\begin{equation}\label{dryuma:eq33}
\left ({\frac {d}{ds}}x(s)\right )U(s)+\left ({\frac {d}{ds}}y(s)
\right )V(s)+\left ({\frac {d}{ds}}z(s)\right )W(s)-1/2\,s=0.
\end{equation}

   Using this relation we find the equation for determination of the function $U(s)$
\begin{equation}\label{dryuma:eq34}
{\frac {d^{2}}{d{s}^{2}}}U(s)+2\,{\frac {\left ({\frac
{d}{ds}}x(s) \right ){c}^{2}{\frac {d}{ds}}U(s)}{\left
({c}^{2}-{\Omega}^{2}\left ( x(s)\right )^{2}\right )x(s)}}+{\frac
{{c}^{4}\left ({\frac {d}{ds}}y( s)\right )^{2}\left
({c}^{2}-3\,{\Omega}^{2}\left (x(s)\right )^{2} \right
)U(s)}{\left (c-\Omega\,x(s)\right )^{3}\left (c+\Omega\,x(s)
\right )^{3}}}+\]\[+{\frac {{c}^{2}}{\left
({c}^{2}-{\Omega}^{2}\left (x(s) \right )^{2}\right )x(s)}} =0.
\end{equation}

    Using the substitution
\[
U(s)=F(s){e^{1/2\,\ln (c-\Omega\,x(s))+1/2\,\ln
(c+\Omega\,x(s))-\ln ( x(s))}}
\]
we can transform the equation (\ref{dryuma:eq34}) into
\begin{equation}\label{dryuma:eq35}
{\frac {d^{2}}{d{s}^{2}}}F(s)-3\,{\frac {{c}^{2}{\Omega}^{2}\left
({c} ^{2}\left ({\frac {d}{ds}}x(s)\right )^{2}+{c}^{2}\left
({\frac {d}{ds }}y(s)\right )^{2}\left (x(s)\right
)^{2}-{\Omega}^{2}\left ({\frac {d }{ds}}x(s)\right )^{2}\left
(x(s)\right )^{2}\right )F(s)}{\left (c- \Omega\,x(s)\right
)^{3}\left (c+\Omega\,x(s)\right )^{3}}}+\]\[+{\frac {{c
}^{2}}{\left (c-\Omega\,x(s)\right )^{3/2}\left
(c+\Omega\,x(s)\right )^{3/2}}} =0. \end{equation}

   Its  solutions is depended from choice of geodesic in  basic space $x(s)$ and
   $y(s)$.

   For example  from the relations
\[
{\frac {d}{ds}}y(s)=\nu\,\left (1-{\frac {{\Omega}^{2}\left (x(s)
\right )^{2}}{{c}^{2}}}\right )\left (x(s)\right )^{-2} ,\quad
{\frac {d}{ds}}x(s)=\sqrt {1+{\frac
{{\nu}^{2}{\Omega}^{2}}{{c}^{2}}}- {\frac {{\nu}^{2}}{\left
(x(s)\right )^{2}}}}
\]
 follows that
 \[
y(s)=const. \quad x(s)=s
 \]
are solutions of geodesics at the parameter $\nu=0$.

  Under these conditions the equation ~(\ref{dryuma:eq35}) takes the form
\begin{equation}\label{dryuma:eq36}
{\frac {d^{2}}{d{s}^{2}}}F(s)-3\,{\frac {F(s){c}^{2}{\Omega}^{2}}{
\left (\Omega\,s-c\right )^{2}\left (\Omega\,s+c\right
)^{2}}}-{\frac {{c}^{2}}{\left (\Omega\,s-c\right )^{3/2}\left
(\Omega\,s+c\right )^{ 3/2}}}=0
\end{equation}
and its solutions are defined by
\begin{equation}\label{dryuma:eq37}
F(s)=1/2\,{\frac {{c}^{2}\sqrt {\Omega\,s-c}\sqrt {\Omega\,s+c}}{{
\Omega}^{2}\left ({s}^{2}{\Omega}^{2}-{c}^{2}\right )}}+{\frac
{{C_1}\,\sqrt {\left (\Omega\,s-c\right )\left (\Omega\,s+c\right
)} \left (\Omega\,s-c\right )}{\Omega\,s+c}}+\]\[+{\frac
{{C_2}\,\sqrt { \left (\Omega\,s-c\right )\left (\Omega\,s+c\right
)}s}{\left (\Omega \,s-c\right )\left (\Omega\,s+c\right )}}
\end{equation}
with two parameters $C_1,C_2$.

  Now with the help of formulae
  (\ref{dryuma:eq33},\ref{dryuma:eq34})
it is possible to find the expressions for coordinates $U(s)$ and
$V(s)$.

   Corresponding  relation $ U=U(V)$ give us the shape of geodesic line $x(y)$
    of basic space in  additional space.

  The generalization and the interpretation of considered solutions will
be done later.

\end{document}